\input{aipcheck}

\documentclass[
    ,final            
  ]
  {aipproc}

\layoutstyle{8x11single}

\begin{document}

\title{Digital radio detection of cosmic rays: achievements, status and perspectives}


\author{Tim Huege}{
  address={Karlsruhe Institute of Technology, Institut f\"ur Kernphysik, Karlsruhe}
}

\begin{abstract}
Over the past decade, radio detection of cosmic rays has matured from small-scale
prototype experiments to installations spanning several km$^2$ with more than a hundred
antennas. The physics of the radio signal is well understood and simulations and measurements
are in good agreement. We have learned how to extract important cosmic ray parameters such as
the geometry of the air shower and the energy of the primary particle from the radio signal,
and have developed very promising approaches to also determine the mass of the primary particles.
At the same time, limitations have become increasingly clear. I review 
the progress made in the past decade and provide a personal view on further
potential for future development. 
\end{abstract}

\maketitle


\section{Introduction}

A decade has passed since the field of radio detection of cosmic ray 
air showers has been revived by the LOPES \citep{FalckeNature2005} and 
CODALEMA \citep{ArdouinBelletoileCharrier2005} experiments. 
Since then, we have progressed from small-scale prototype installations 
to arrays with instrumented areas of several km$^{2}$. A fundamental 
breakthrough has been achieved with a detailed understanding of the radio emission 
mechanisms. On the basis of this understanding, analysis 
procedures have been developed to extract the air shower 
characteristics of interest, in particular the arrival direction, 
primary particle energy and depth of shower maximum. In this article I will give a concise overview of the achievements of 
the past decade, focusing in particular on the most recent results,
and then give a short outlook on the future application potential of the
technique.

\section{Achievements of the last decade}

Major milestones on the way towards a full exploitation of the radio 
detection technique have been reached in the past decade. The 
following represents an incomplete selection, focused in particular on 
recent results which have preferably been published in peer-reviewed journal 
articles.

\subsection{Radio emission physics}

\begin{figure}[htb]
  \includegraphics[width=0.23\textwidth]{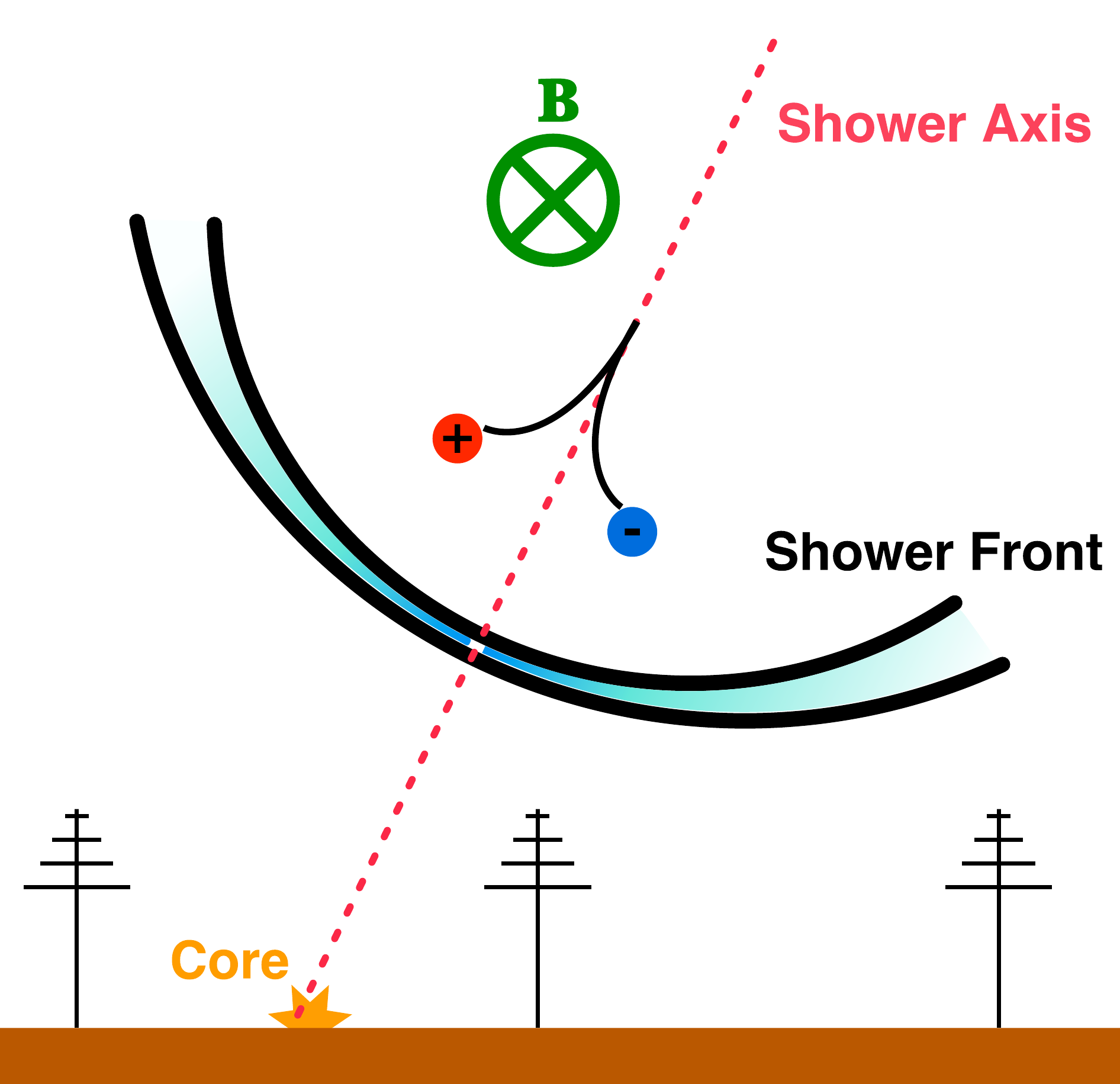}
  \includegraphics[width=0.23\textwidth]{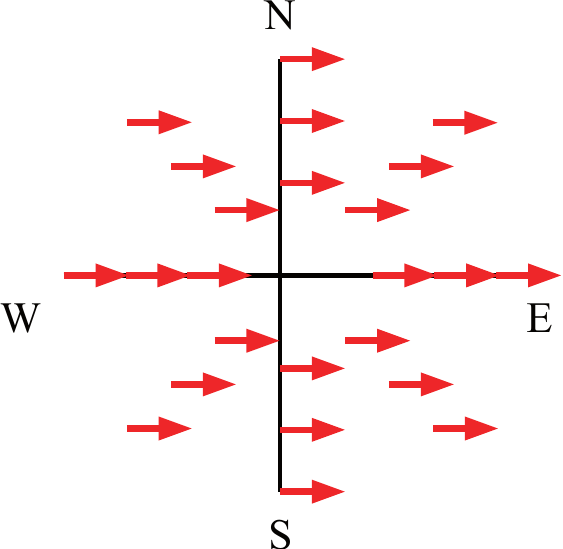}
  \hspace{0.05\textwidth}
  \includegraphics[width=0.23\textwidth]{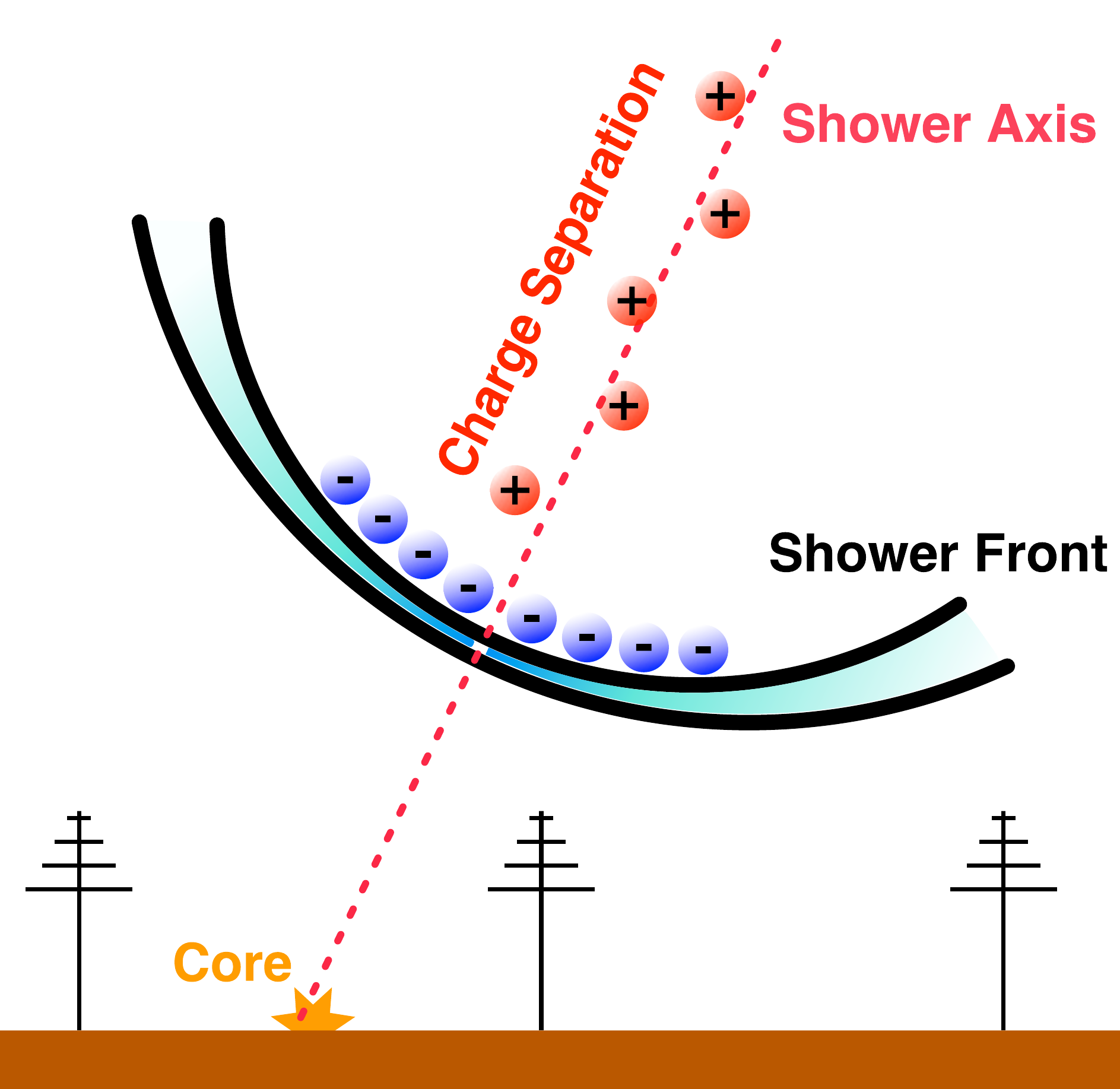}
  \includegraphics[width=0.23\textwidth]{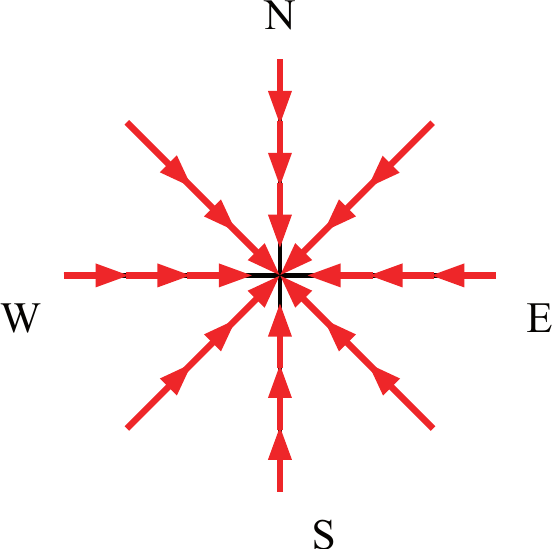}
  \caption{The radio emission from air showers can be interpreted as a 
  superposition of geomagnetic (left) 
  and charge excess (right) emission with specific polarization 
  patterns. Diagrams are from \citep{SchoorlemmerThesis2012} and K.D. 
  de Vries.}\label{fig:emission}
\end{figure}

\begin{figure}[htb]
  \includegraphics[width=0.55\textwidth]{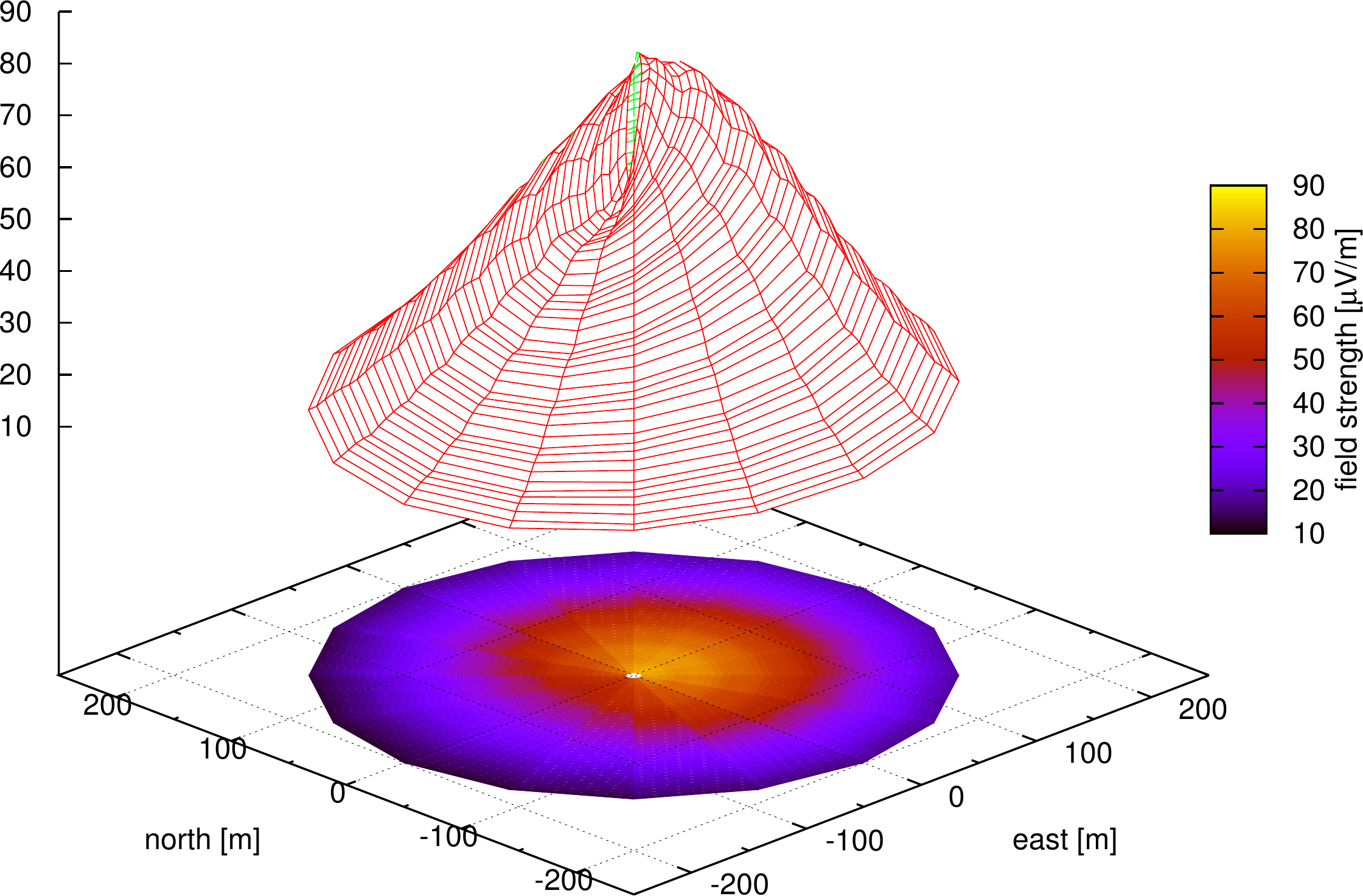}
  \caption{Superposition of the geomagnetic and charge excess 
  contributions produces a pronounced asymmetry in the footprint of 
  the radio emission, as here shown for a CoREAS simulation \citep{HuegeARENA2012a}.}\label{fig:asymmetry}
\end{figure}

Several approaches have been followed to achieve a detailed 
understanding of the physics of the radio emission from extensive air 
showers. It has been very fruitful to approach the problem from two 
perspectives: on the one hand, microscopic  Monte Carlo 
simulations \citep{HuegeARENA2012a,AlvarezMunizCarvalhoZas2012,MarinRevenu2012} have been developed which 
calculate the radio emission based on air shower simulations and classical 
electrodynamics without any free parameters; on the other hand, 
macroscopic models \citep{DeVriesVanDenBergScholten2010,WernerDeVriesScholten2012}, have been used to calculate the 
emission by superposing ``emission mechanisms''. From the combination 
of all these works, we have learned that the radio emission can be 
interpreted as a superposition of geomagnetic radiation from 
time-varying transverse currents \citep{KahnLerche1966} and the 
Askaryan effect \citep{Askaryan1962a} induced by the time-varying net 
charge excess in the shower (Fig.\ \ref{fig:emission}). Both mechanisms have peculiar polarization 
characteristics. Their superposition leads to a pronounced asymmetry in the 
radio emission footprint on the ground, as shown in Fig.\ 
\ref{fig:asymmetry}. The 
emission from individual electrons and positrons in the shower adds up 
coherently, especially at frequencies below 100~MHz. The refractive 
index gradient in the atmosphere leads to a relativistic 
time-compression of the emitted radio pulses for observers located 
near the Cherenkov angle, so that at these locations emission can be 
observed up to GHz frequencies \citep{SmidaWernerEngel2014}.

\begin{figure}[h!tb]
  \includegraphics[width=0.3\textwidth]{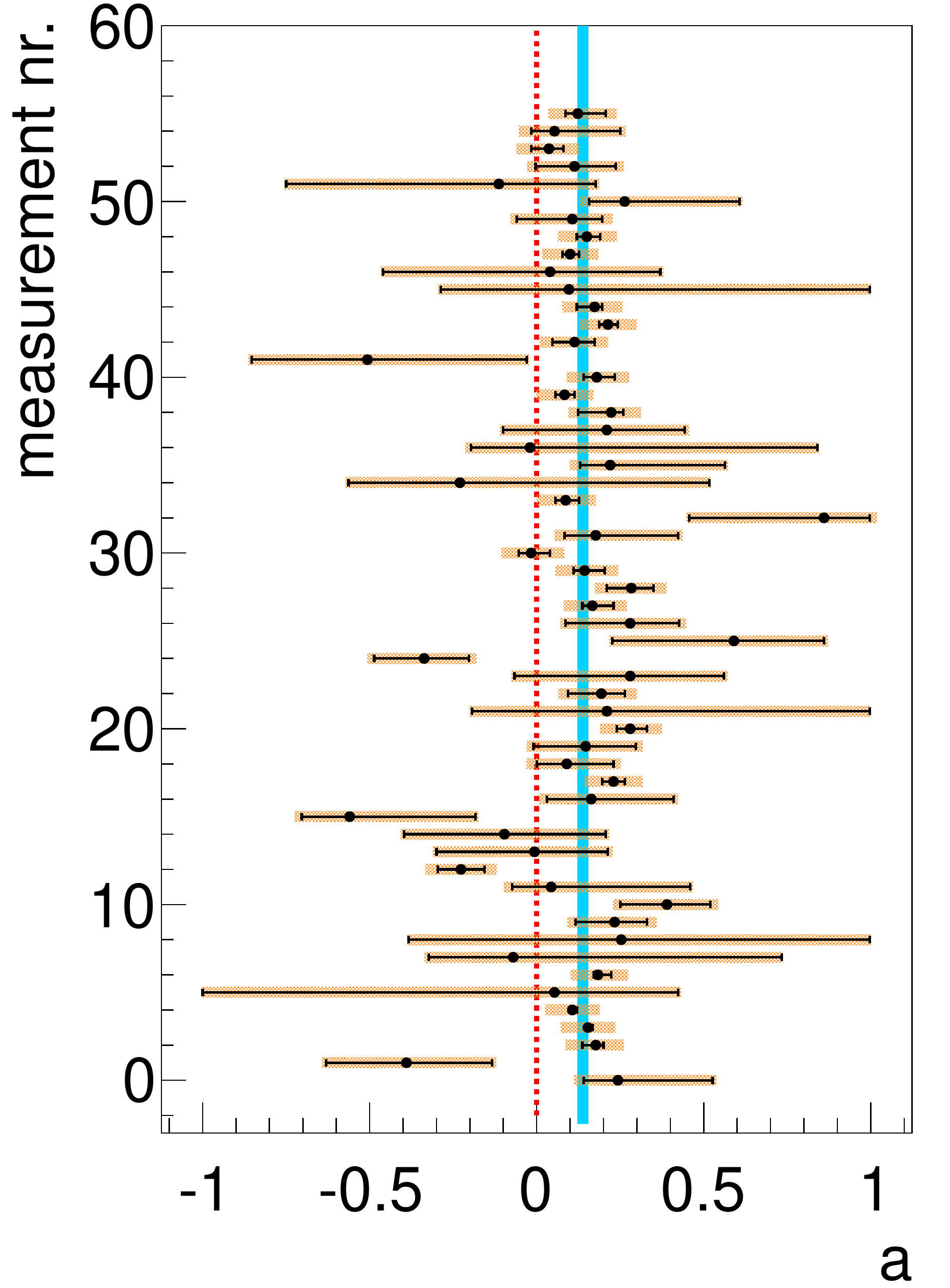}
  \includegraphics[width=0.5\textwidth]{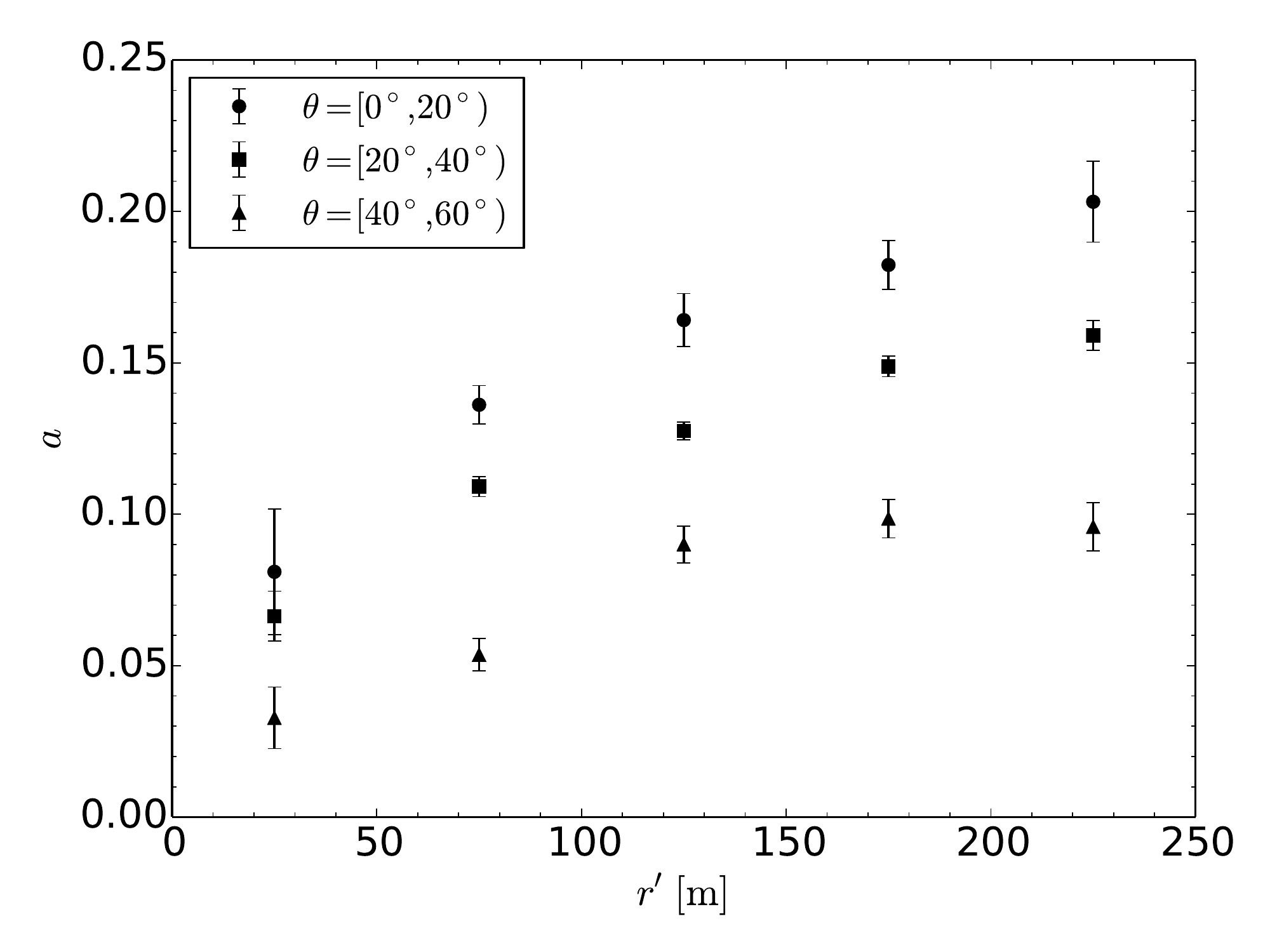}
  \caption{With AERA, the relative contribution of an emission 
  component with radially aligned electric field vectors, as expected 
  for the charge excess emission, was quantified in individual 
  measurements (left). The average relative strength was measured to be 
  14\% \citep{AERAPolarization2014}. 
  With LOFAR, it could be shown that the relative strength is a 
  function of zenith angle and lateral distance (right) 
  \citep{LOFARChargeExcess2014}.}\label{fig:chargeexcess}
\end{figure}

\subsection{Strength of the charge excess contribution}

While the geomagnetic emission mechanism has been identified already 
in the historical works \citep{Allan1971}, the charge excess contribution was 
first established by the modern experiments. An analysis of CODALEMA 
data first established the asymmetry of the radio footprint, a 
signature of the superposition of both mechanisms \citep{CODALEMACoreShift}. With  AERA 
the first quantitative measurement of an emission contribution 
with radially aligned electric field vectors was made \citep{SchroederAERAIcrc2013}. The relative 
strength of this emission was quantified with an average value of 14\% 
(Fig.\ \ref{fig:chargeexcess} left). With LOFAR \citep{NellesIcrc2013} it has now been shown that this value is in 
fact not a constant but varies with lateral distance and zenith angle 
(Fig.\ \ref{fig:chargeexcess} right), 
as is also expected from simulations. Readers should keep in mind that 
for different experiments with different geomagnetic field 
configurations, the values are expected to be different. It is important to know the 
strength of this relative contribution as well as possible so that it can be corrected for 
in analyses procedures.

\subsection{Energy reconstruction}

It has long been known that the radio electric field strength 
scales linearly with the primary particle energy, as it should for 
coherent radiation. Exploiting geometrical influences on the emission, 
the influence of shower-to-shower-fluctuations can be minimized and 
simulations predict an intrinsic energy resolution of radio 
measurements of well below 10\% \citep{HuegeUlrichEngel2008}.

\begin{figure}[htb]
  \includegraphics[clip=true,trim=0 0 90 400,width=0.5\textwidth]{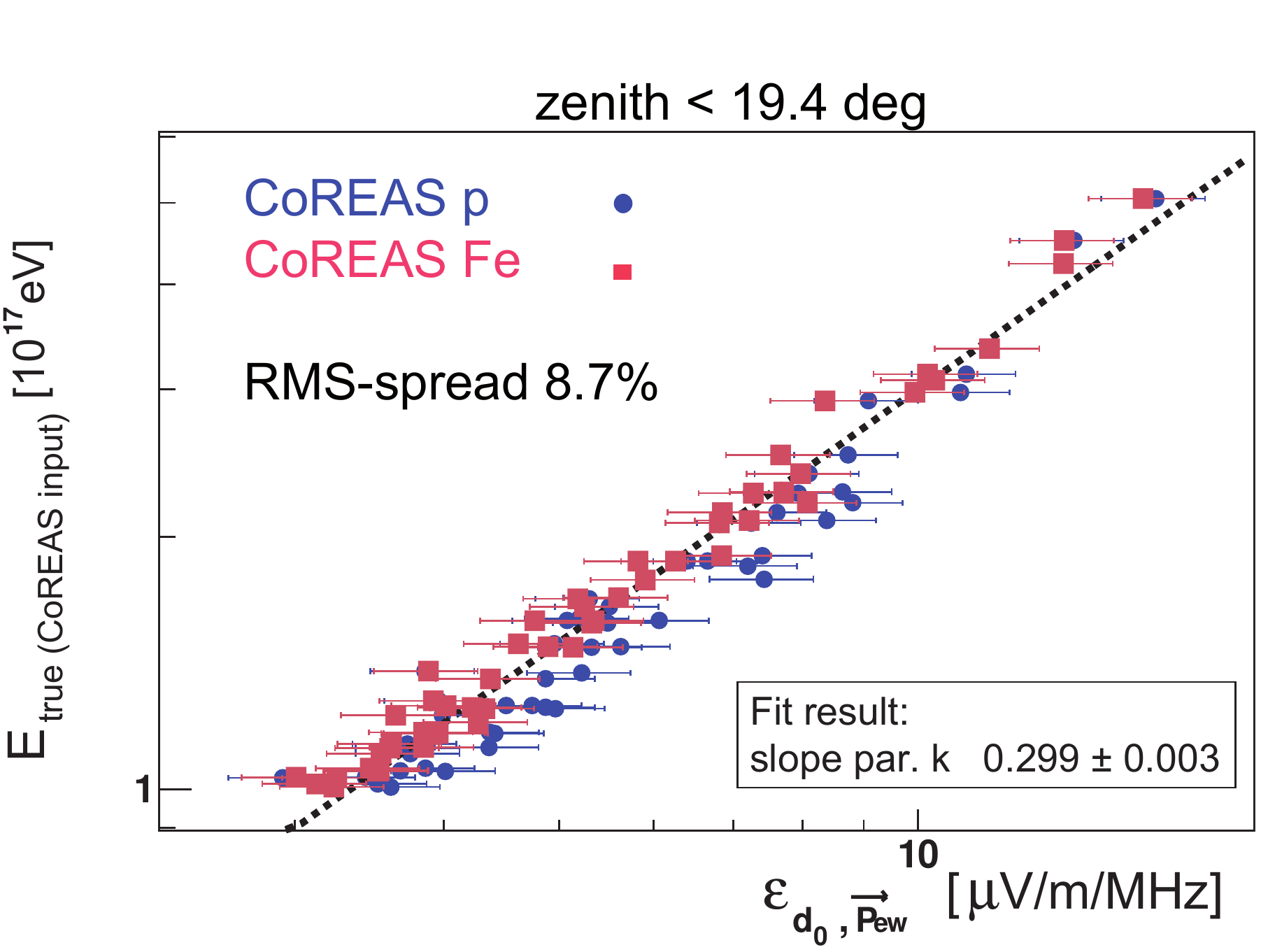}
  \includegraphics[clip=true,trim=0 0 90 400,width=0.5\textwidth]{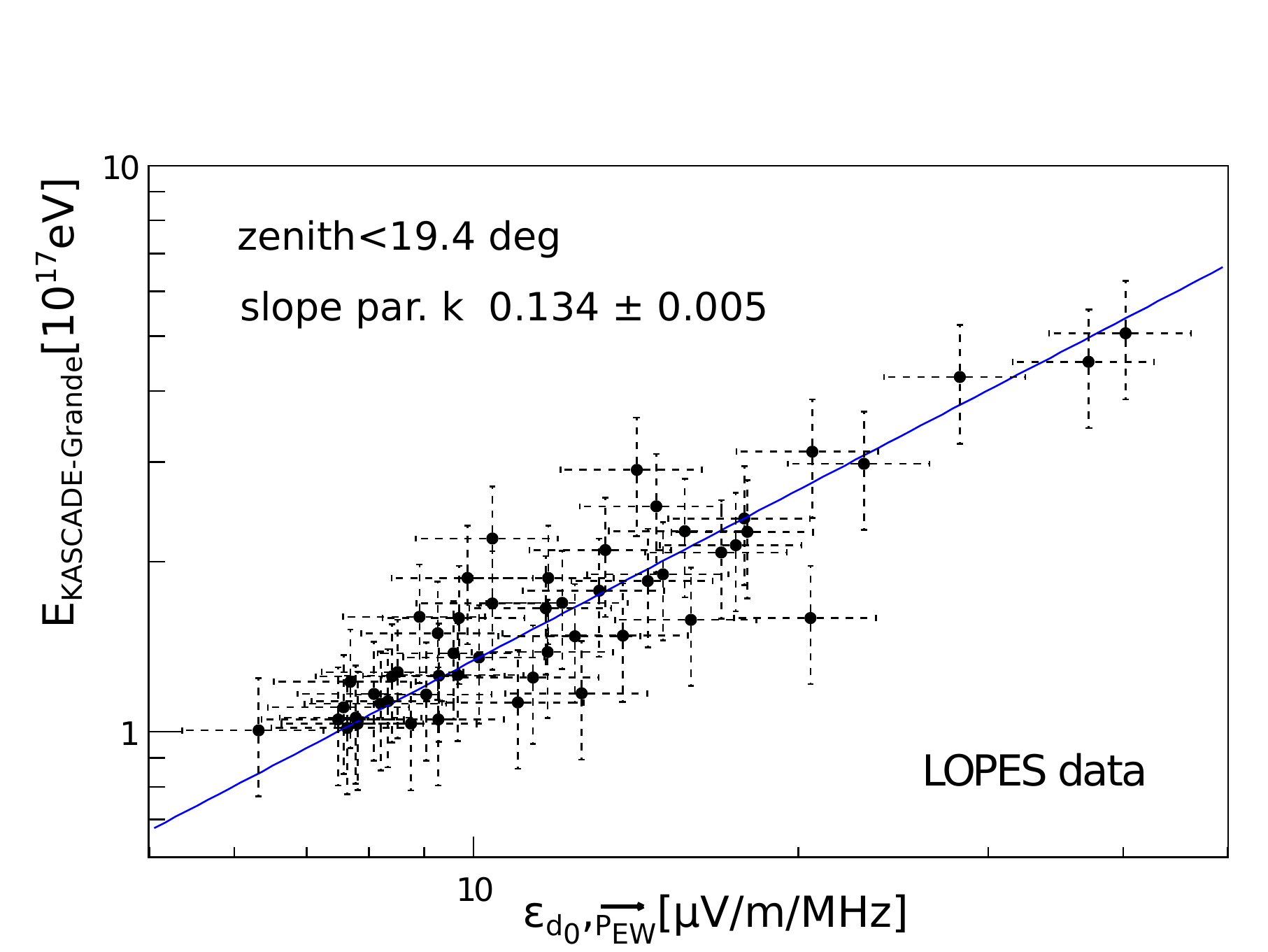}
  \caption{Correlation between radio amplitude and Monte Carlo true 
  energy as predicted with CoREAS simulations for LOPES data (left) 
  and actual linear correlation measured with LOPES (right) 
  \citep{ApelArteagaBaehren2014}.}\label{fig:energy}
\end{figure}

Many experiments have reported correlations of particle energy and radio 
amplitudes, but the only quantitative publication in a peer-reviewed 
journal so far has been made 
by LOPES \citep{ApelArteagaBaehren2014}. This analysis confirmed with 
modern simulations that 
intrinsically an energy resolution of 
better than 10\% should be achievable (Fig.\ \ref{fig:energy} left). 
Measured LOPES data confirm the predicted linear scaling and achieve a 
combined energy resolution 
of KASCADE-Grande and LOPES of $\approx$~20\% (Fig.\ \ref{fig:energy} 
right). This is remarkable as 
the KASCADE-Grande energy resolution alone is approximately 20\%. Work 
on energy determination by Tunka-Rex \citep{SchroederTRexIcrc2013} and 
AERA is ongoing, and initial 
results having been shown at this conference. An interesting point 
will be to compare the absolute energy scales derived with the 
different experiments in light of the challenge of absolute 
calibration of the detectors.

Precise calorimetric energy measurements with radio antennas can be 
very valuable to cross-check the energy scales of other techniques 
such as surface particle detectors and fluorescence telescopes. The 
intrinsic high energy resolution, the sensitivity to the 
electromagnetic component of the air shower alone, and the fact that 
the radio emission can be calculated from first principles and does 
not suffer attenuation in the atmosphere makes this approach very 
promising for future studies.

\clearpage

\subsection{Xmax reconstruction}

Radio emission from air showers contains information on the distance 
of the radio source and thus the depth of shower maximum \citep{HuegeUlrichEngel2008}, Xmax, the 
primary mass-sensitive parameter used in air shower physics. First 
experimental proof that the lateral distribution function of radio 
signals is sensitive to the shower 
evolution, in this case probed with muon pseudorapidities measured by 
KASCADE-Grande, was published by the LOPES experiment \citep{ApelArteagaBaehren2012c}.

\begin{figure}[htb]
  \includegraphics[clip=true,trim=0 0 90 350,width=0.4\textwidth]{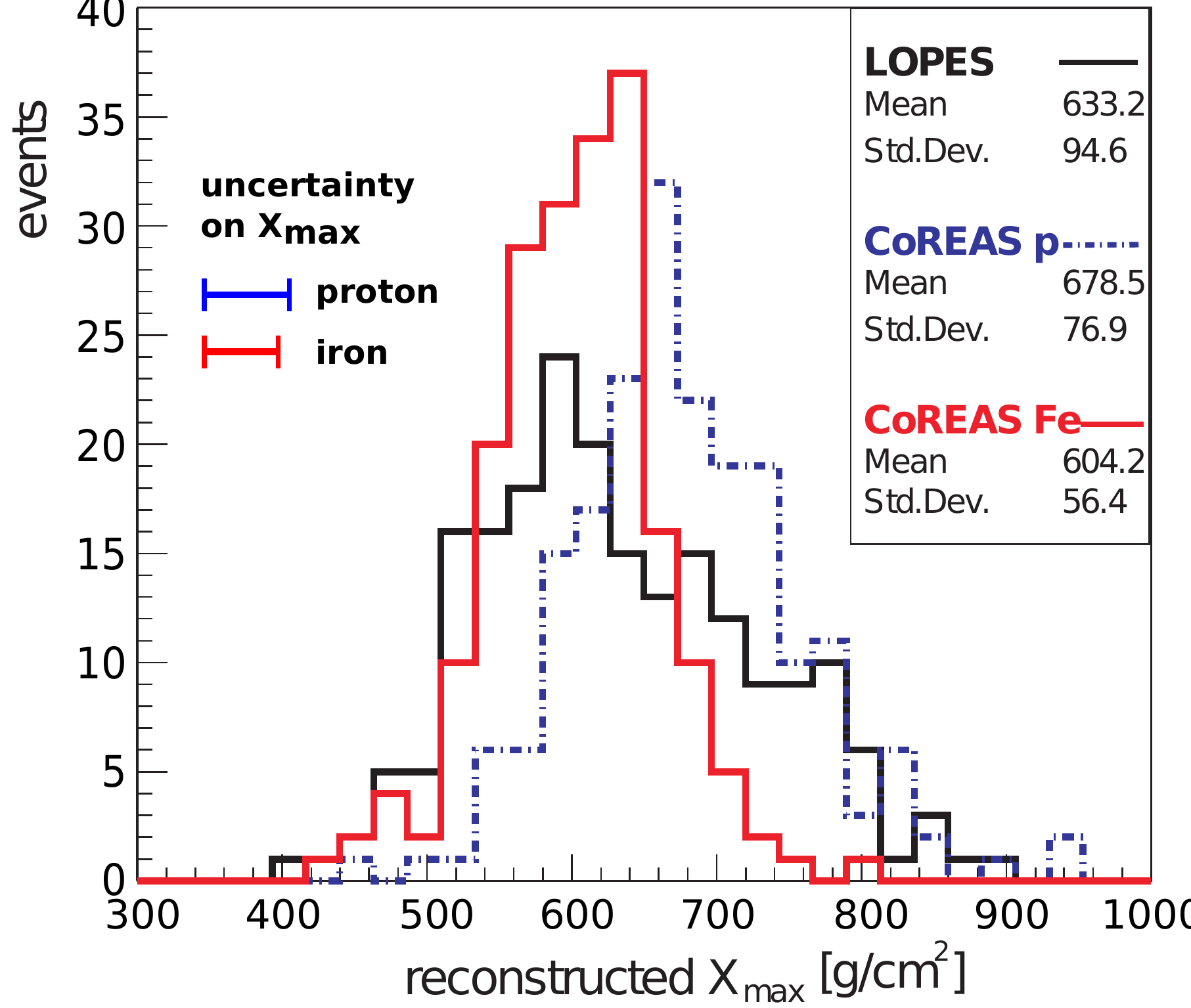}
  \includegraphics[clip=true,trim=0 0 0 460,width=0.7\textwidth]{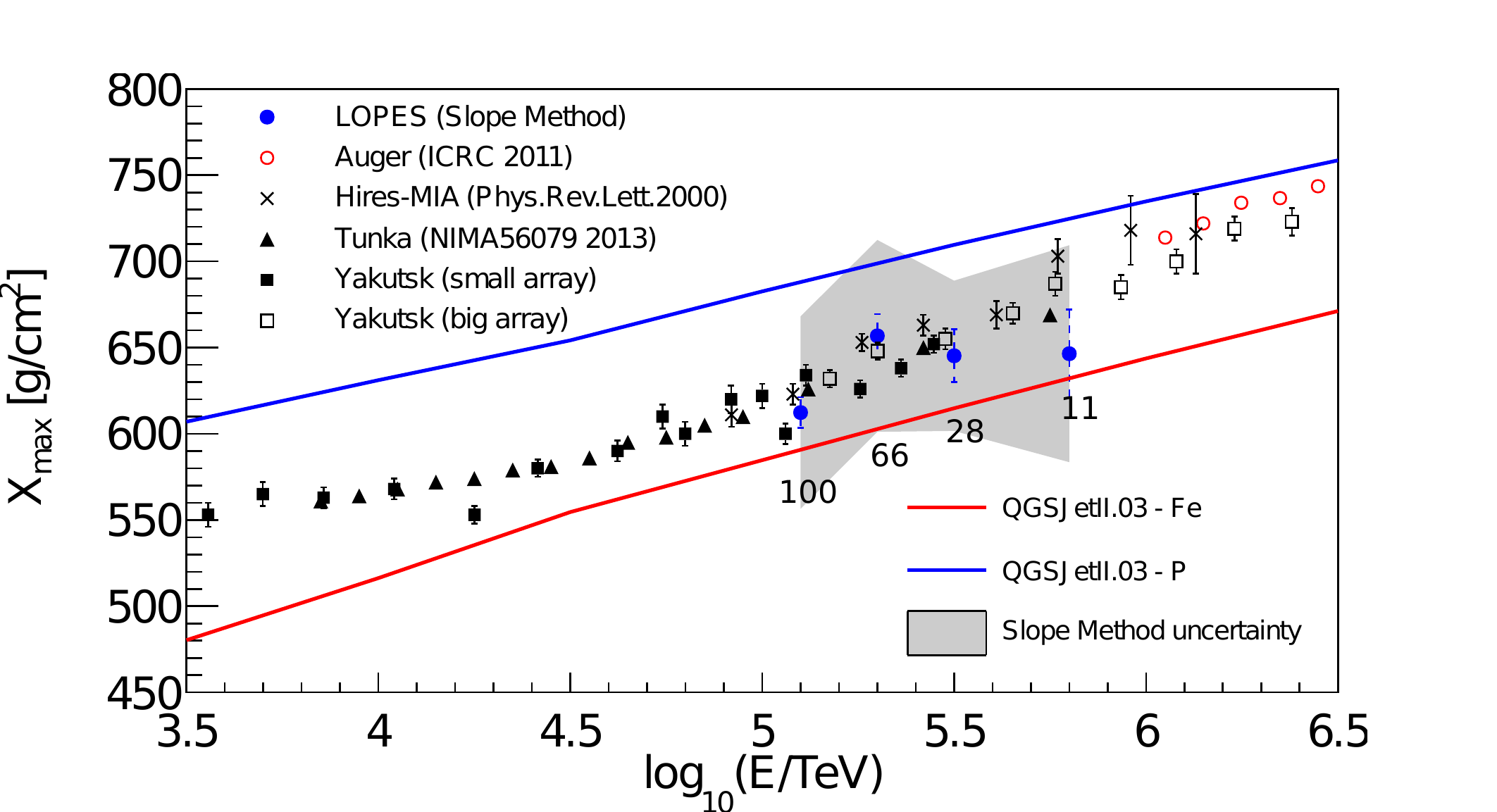}
  \caption{From the slope of the radio lateral distribution functions, 
  an Xmax measurement for each individual showers can be derived from 
  LOPES data in combination with CoREAS simulations (left). The 
  resulting mean Xmax values as a function of energy are consistent 
  with measurements from other experiments (right). Both results are from 
  \citep{ApelArteagaBaehren2014}.}\label{fig:xmaxlopes}
\end{figure}

A quantitative analysis of LOPES data in direct comparison with CoREAS 
simulations showed that from the measured lateral slopes of the radio 
signals, Xmax values can be reconstructed which are consistent with 
the expectation for a mass composition between the extremes of pure proton and iron primaries (Fig.\ 
\ref{fig:xmaxlopes} left) and measurements from established 
experiments with Xmax sensitivity (Fig.\ \ref{fig:xmaxlopes} right). 
The method applied to LOPES data was kept fairly simple, as the data 
quality was a limiting factor and no independent reference Xmax 
information could be used for comparison. The resulting method 
uncertainty was 50~g~cm${^{-2}}$ and the total systematic uncertainty was 
estimated to at most 90~g~cm$^{-2}$, which is not competitive with 
measurements by established techniques such as Fluorescence detection, 
which reaches a resolution of 20~g~cm$^{-2}$.

\begin{figure}[htb]
  \includegraphics[width=0.34\textwidth]{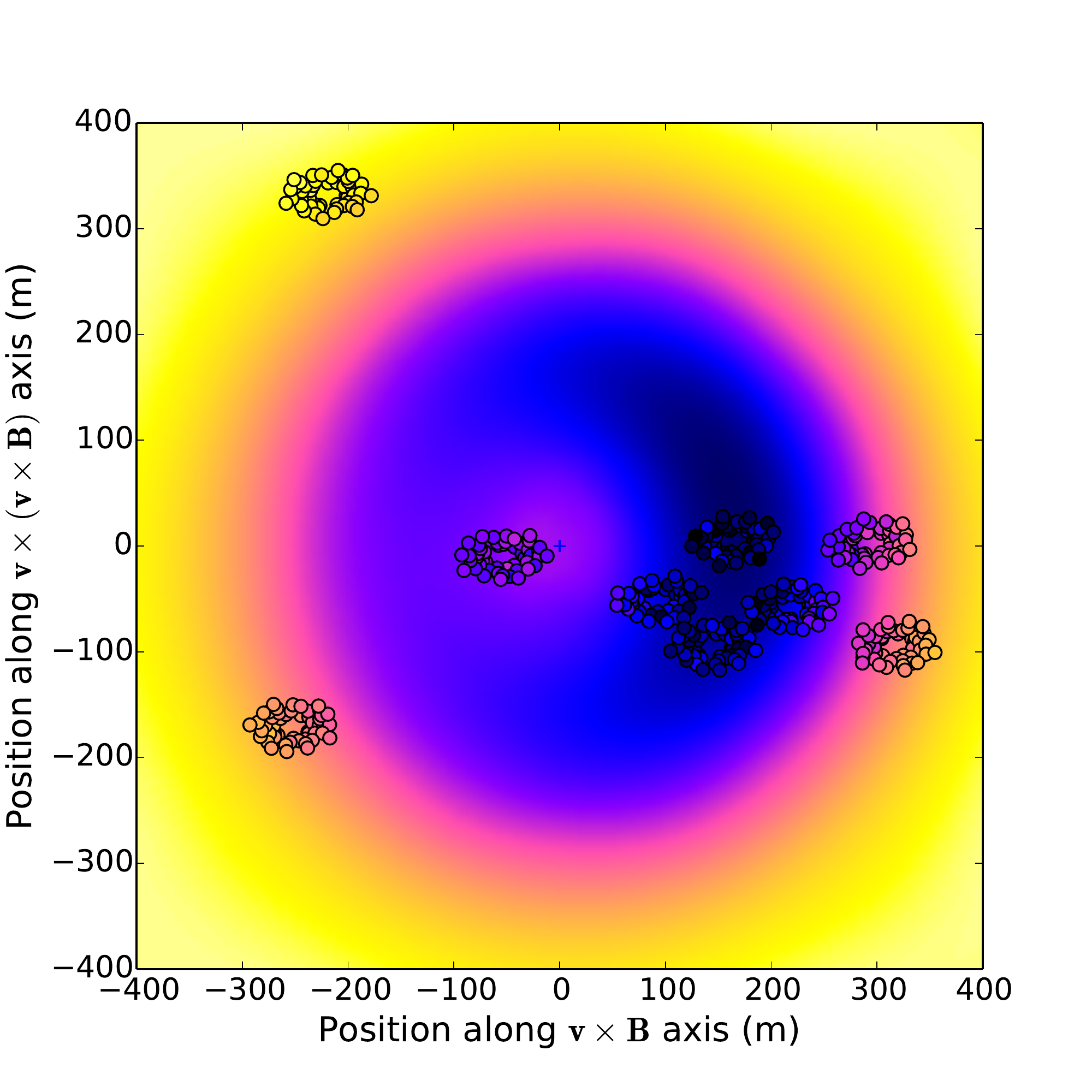}
  \includegraphics[clip=true,trim=50 0 50 0,width=0.38\textwidth]{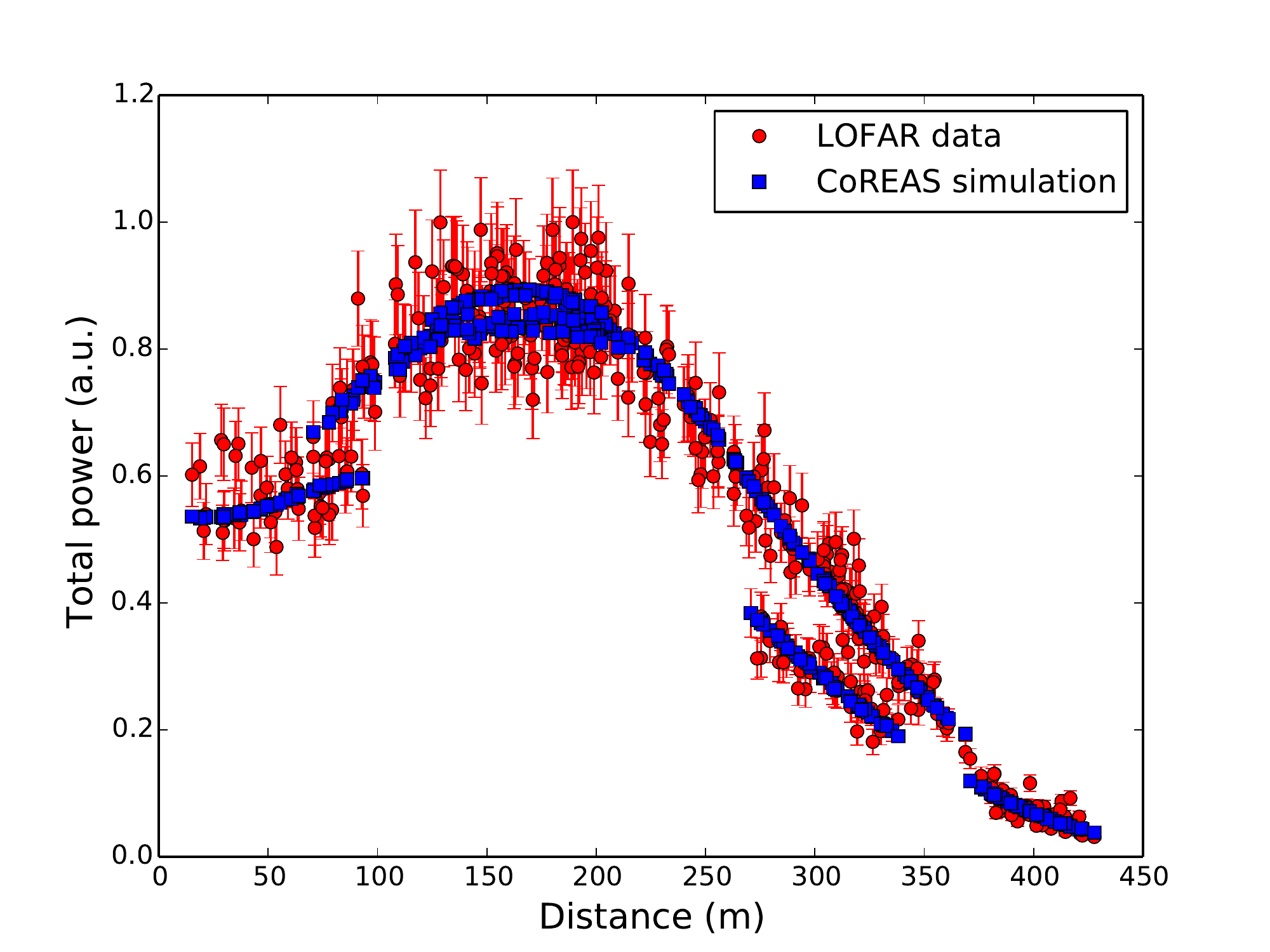}
  \includegraphics[width=0.34\textwidth]{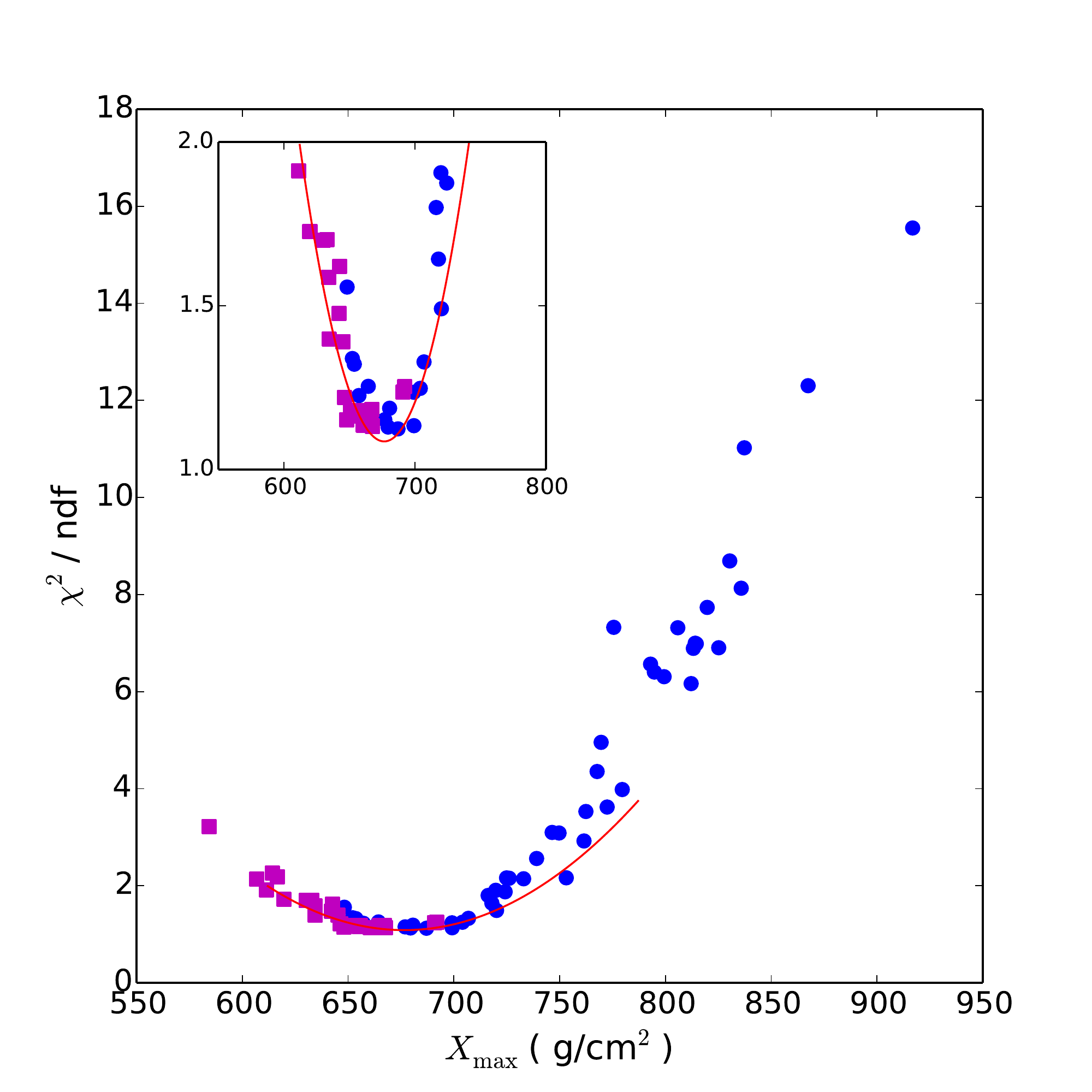}
  \caption{Comparison between a LOFAR measurement of an air shower 
  (left, circles) and a CoREAS simulation (left, background color). A 
  one-dimensional projection (middle) illustrates the impressive 
  agreement and the pronounced structure due to Cherenkov-like time 
  compression and asymmetries. The dominating parameter governing fit quality 
  is Xmax, which can in turn be determined with high resolution from 
  comparison with multiple simulations
  (right). All diagrams are from \citep{LOFARXmax2014}.}\label{fig:xmaxlofar}
\end{figure}

\clearpage

LOFAR has recently demonstrated, however, that if a detailed 
measurement of the complex two-dimensional radio footprint with all 
its asymmetries and structure is available, Xmax can be determined 
with a resolution of less than 20~g~cm$^{-2}$. This is done by 
simulating radio emission profiles and comparing them with 
measurements (Fig. \ref{fig:xmaxlofar} left and middle) until the best 
possible agreement has been found. The dominating parameter governing fit quality is the depth of 
shower maximum, which can in turn be read off from the comparison with 
simulations (Fig. \ref{fig:xmaxlofar} right).

\begin{figure}
  \includegraphics[width=0.5\textwidth]{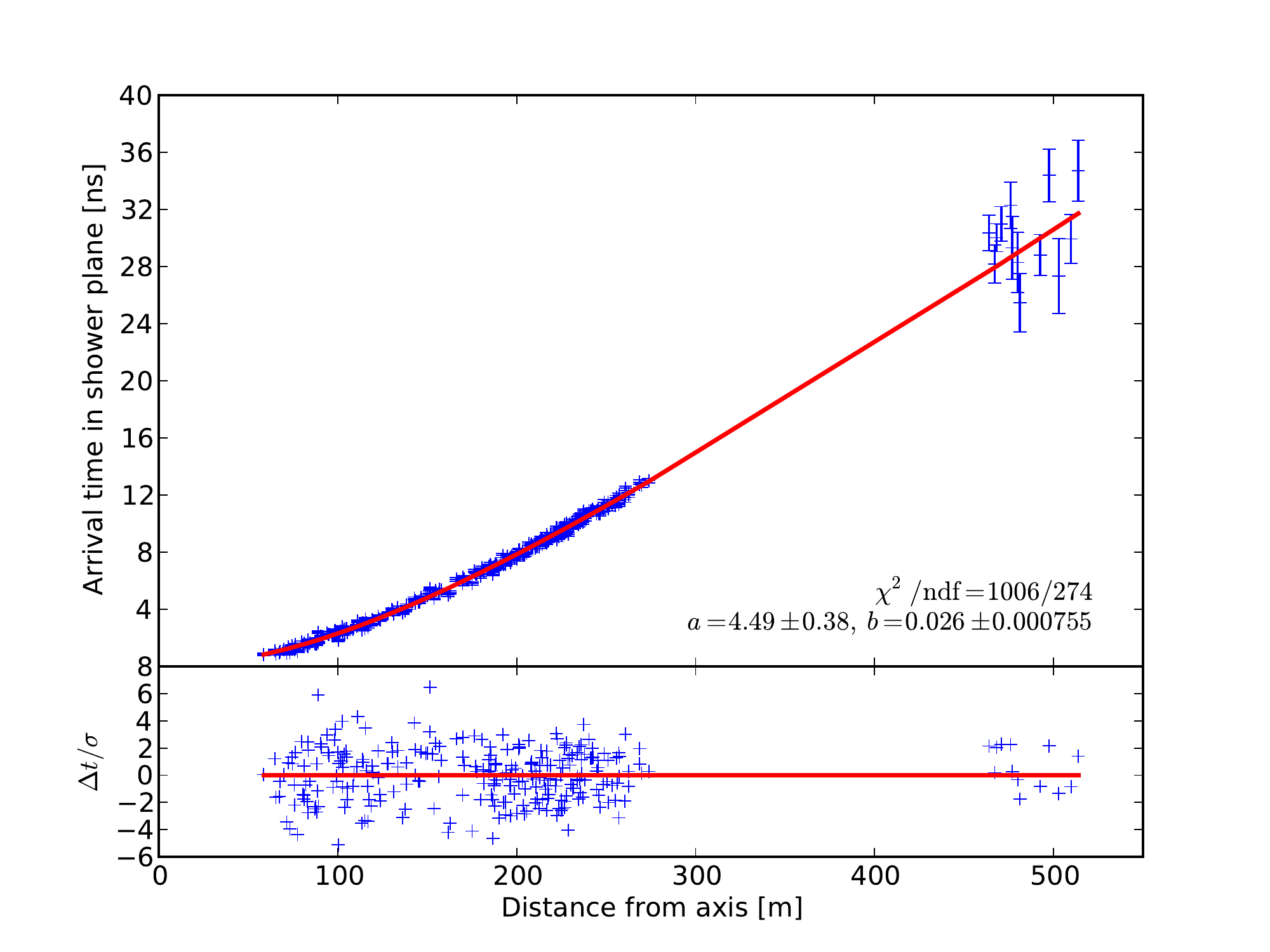}
  \includegraphics[clip=true,trim=0 0 0 350,width=0.6\textwidth]{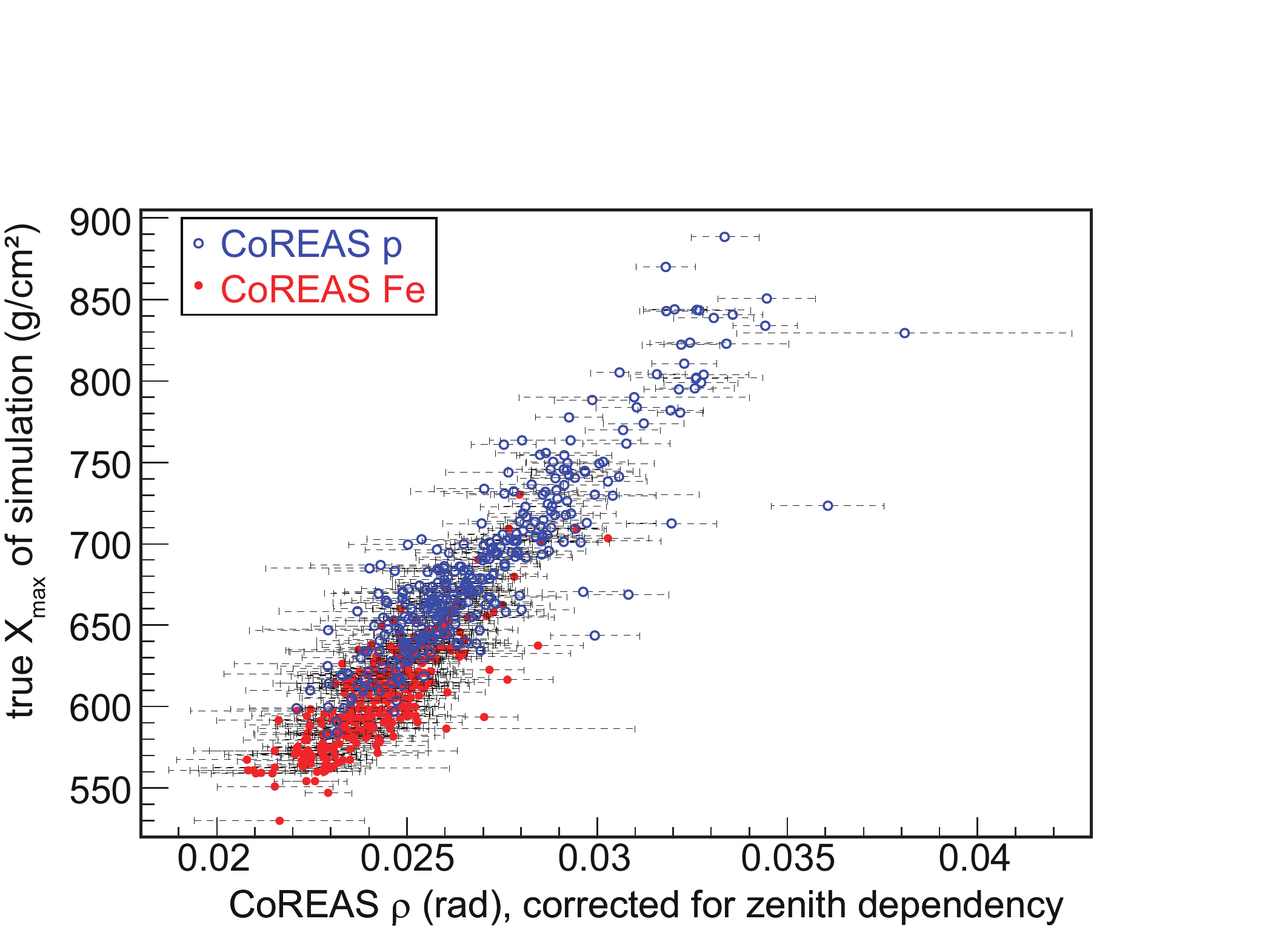}
  \caption{LOFAR has experimentally confirmed that the radio wavefront 
  can be best described by a hyperbola (left) \citep{LOFARWavefront2014}. In a LOPES analysis it 
  has been shown how the opening angle of the asymptotic cone of the 
  hyperbola can be used to determine Xmax (right) \citep{LOPESWavefront2014}.}\label{fig:wavefront}
\end{figure}

In addition to exploiting the amplitude distribution of the radio 
signal, it has long been predicted that also the radio 
emission wavefront can be used to determine the source distance and 
thus Xmax. LOPES had predicted that a hyperbolic wavefront describes 
the arrival times best, which has now been shown very clearly by LOFAR 
(Fig.\ \ref{fig:wavefront} left). After fixing an offset parameter, 
the opening angle of the asymptotic cone of the hyperbola can be used 
to determine Xmax directly, as has been shown in a LOPES analysis on 
the basis of CoREAS simulations (Fig.\ \ref{fig:wavefront} right). The 
achievable precision is predicted to be approximately 30~g~cm$^{-2}$.

Studies at AERA and Tunka-Rex, where independent Xmax measurements are 
available for comparison with the radio results, will provide  
experimental verification of these reconstruction approaches soon. The most 
promising strategy will of course be to combine these two methods in a 
joint analysis, possibly complemented by further observables such as 
the pulse shape or frequency spectral index.


\section{Future potential}

Many successes have already been achieved, and with the detailed 
understanding of the radio emission physics, the potential for the 
future can be better evaluated than it could be a number of years ago. 
I stress that the following is my personal view, which might not be 
shared by everybody else.

\begin{figure}
  \includegraphics[width=0.4\textwidth]{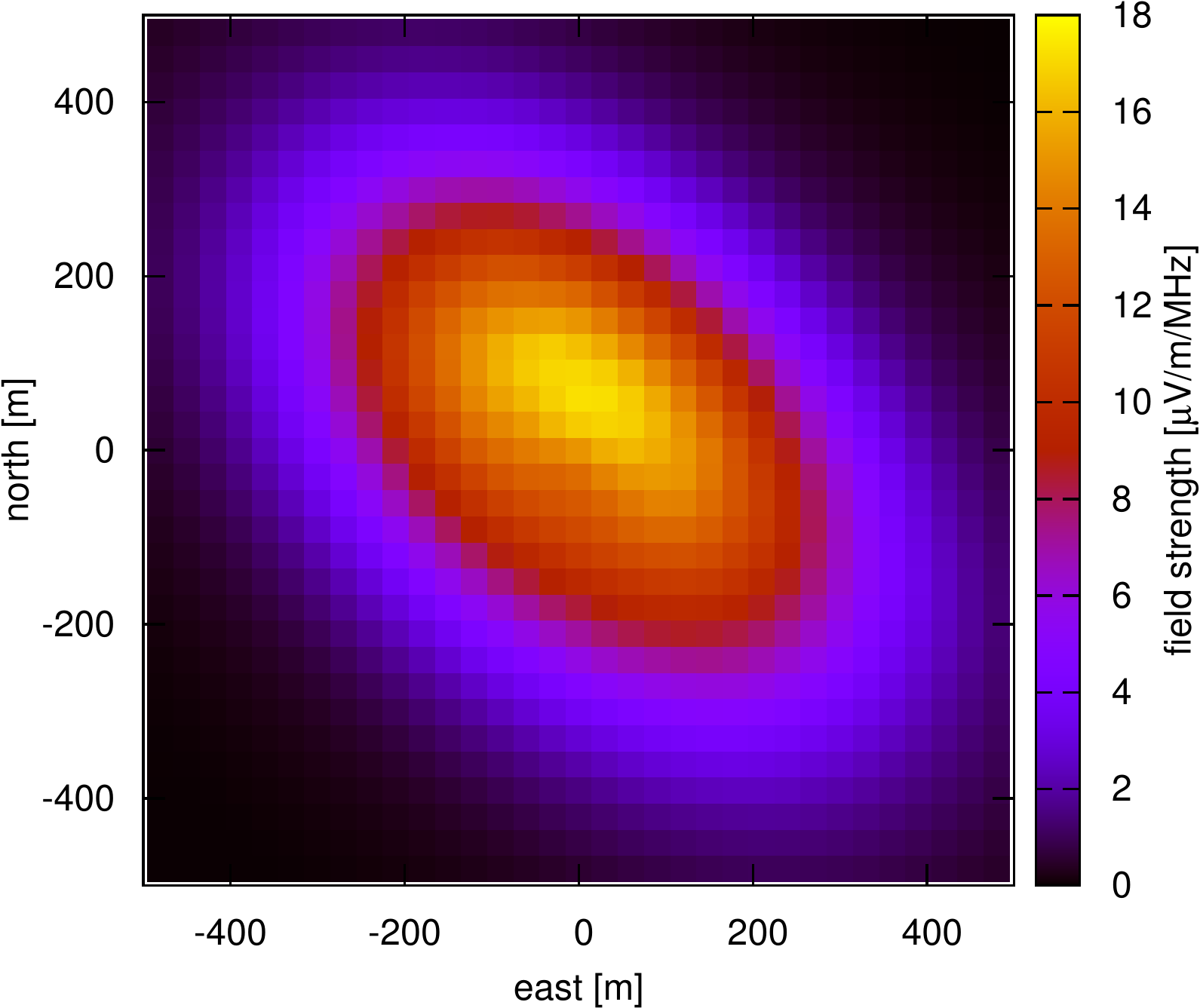}
  \includegraphics[width=0.4\textwidth]{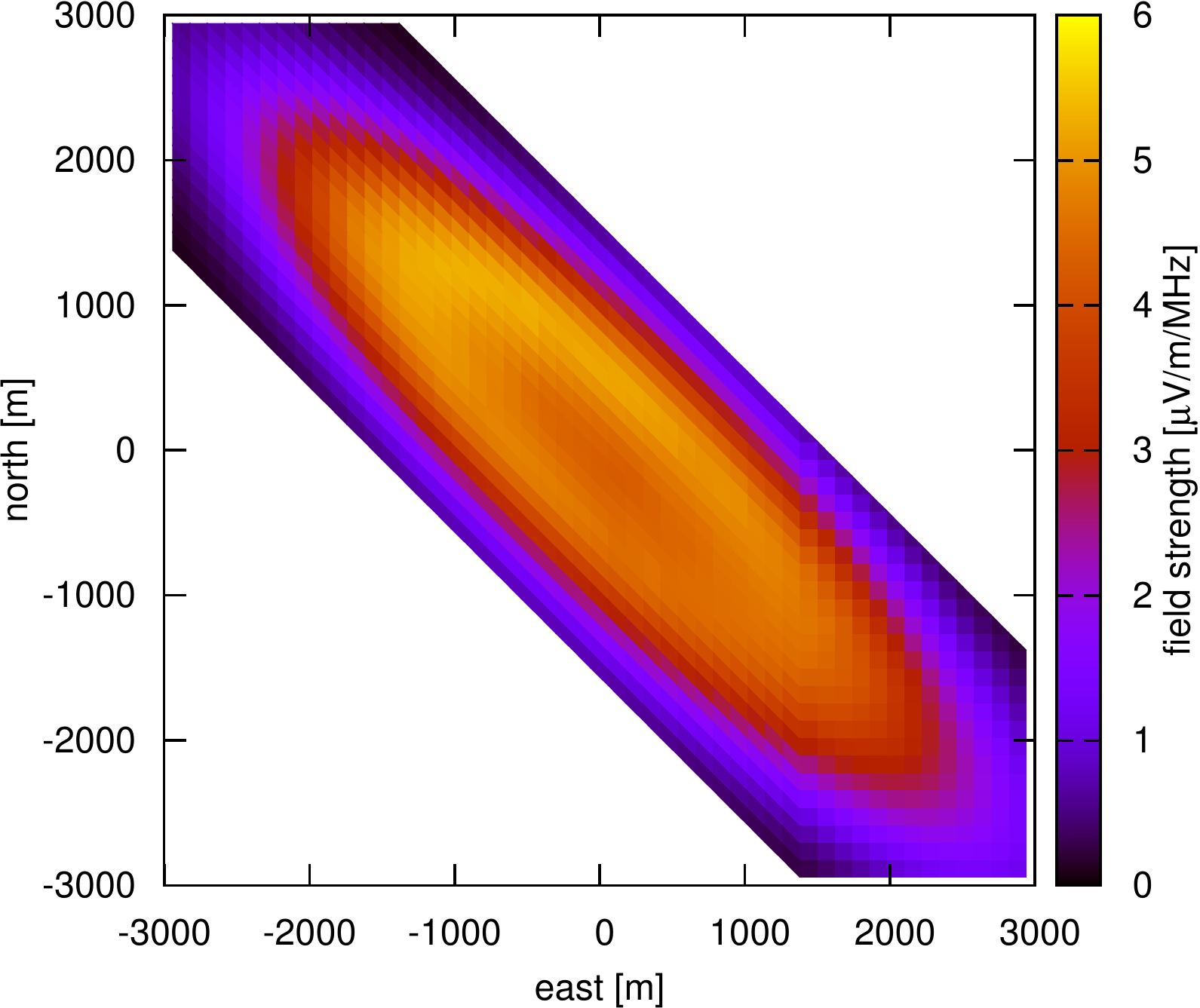}
  \caption{Simulated footprints of radio emission for an air shower 
  with 50$^{\circ}$ (left) and 75$^{\circ}$ (right) zenith angles as 
  predicted with CoREAS simulations \citep{HuegeIcrc2013CoREAS}.}\label{fig:footprintsize}
\end{figure}

Two complementary paths are currently being followed in the field. One 
is to try to cover larger and larger areas. AERA currently covers 
6~km$^{2}$ with 124 detector stations. The original hope was that areas as 
large as the one of the Pierre Auger Observatory with its 
3000~km$^{2}$ could soon be covered with radio antennas as an 
additional detector. It has become clear by now, however, that the 
radio emission footprint for air showers with zenith angles below 
60$^{\circ}$ only has a size of a few hundred meters in radius (Fig. 
\ref{fig:footprintsize}), independent of the primary particle energy. To be 
able to extract Xmax information from radio signals alone it seems that an antenna 
spacing significantly larger than $\approx$~300~m is insufficient. This means that 
large areas would need instrumentation with a very high number of antennas. 
The current concepts are not scalable to arrays with thousands of 
antennas. Considerable research \& development is needed if one wants to achieve this 
goal. Inclined air showers, however, produce very large radio footprints 
because the radio source moves far away (Fig.\ \ref{fig:footprintsize} 
right). This is not a pure projection 
effect and was realized early on \citep{GoussetRavelRoy2004}. While the 
Xmax sensitivity in the radio signal diminishes for inclined air 
showers, coincident detection with radio antennas and particle detectors can yield 
valuable composition information via the ratio of the electromagnetic 
and muonic components in the air shower.

\begin{figure}
\includegraphics[clip=true,trim=80 28 90 35, width=0.32\textwidth]{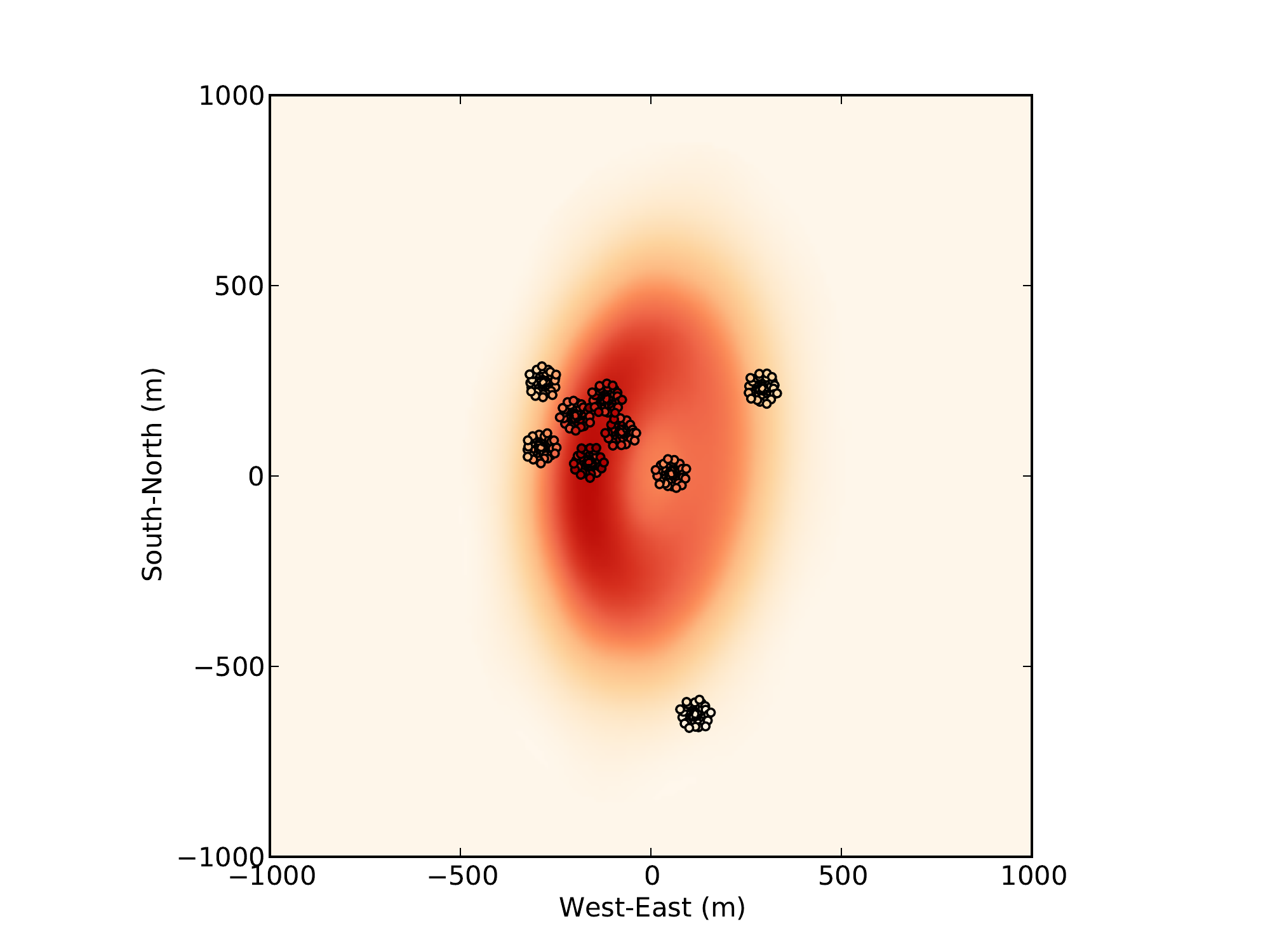}
\includegraphics[clip=true,trim=80 28 90 35, width=0.32\textwidth]{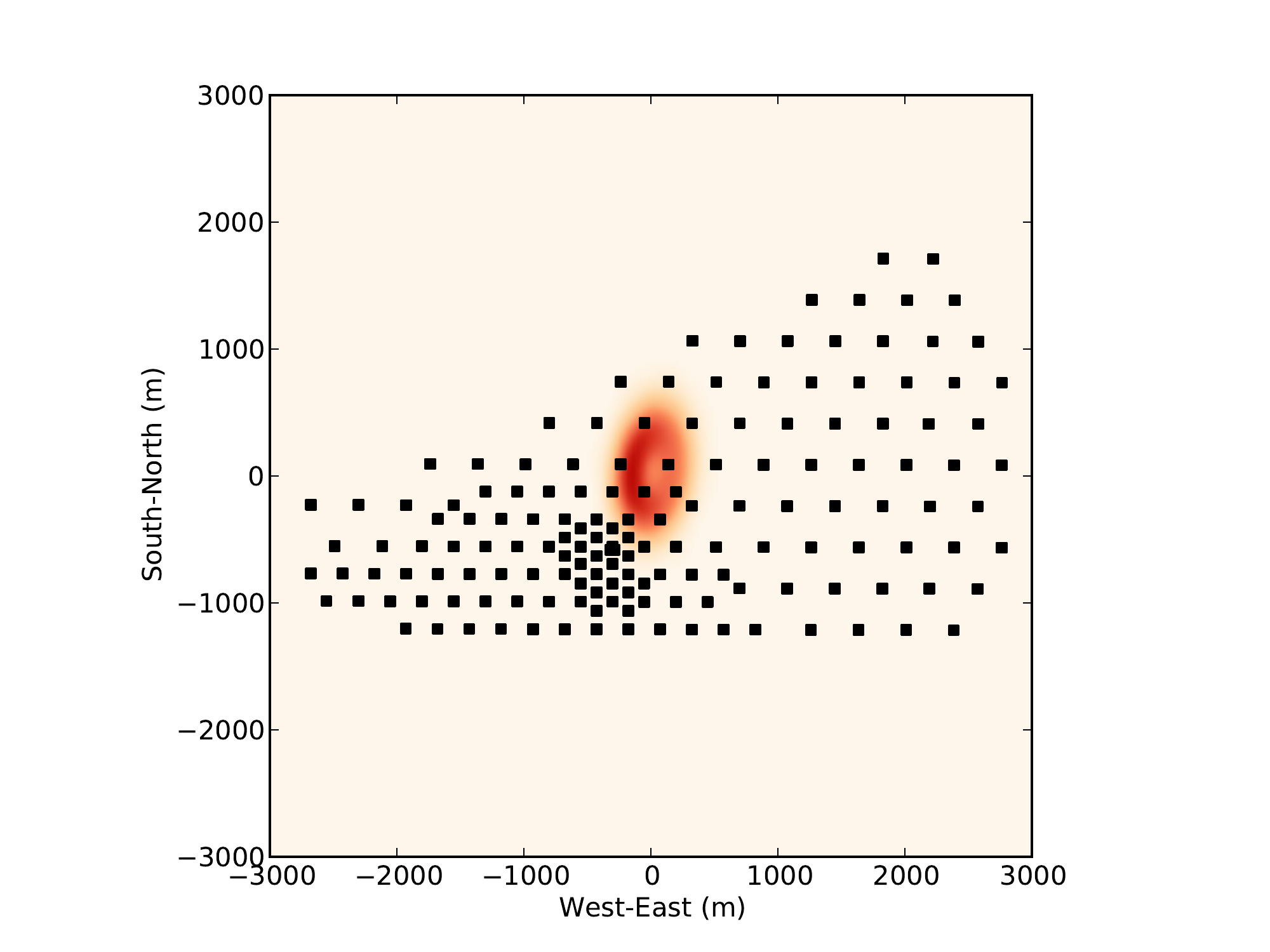}
\includegraphics[clip=true,trim=80 28 90 35, width=0.32\textwidth]{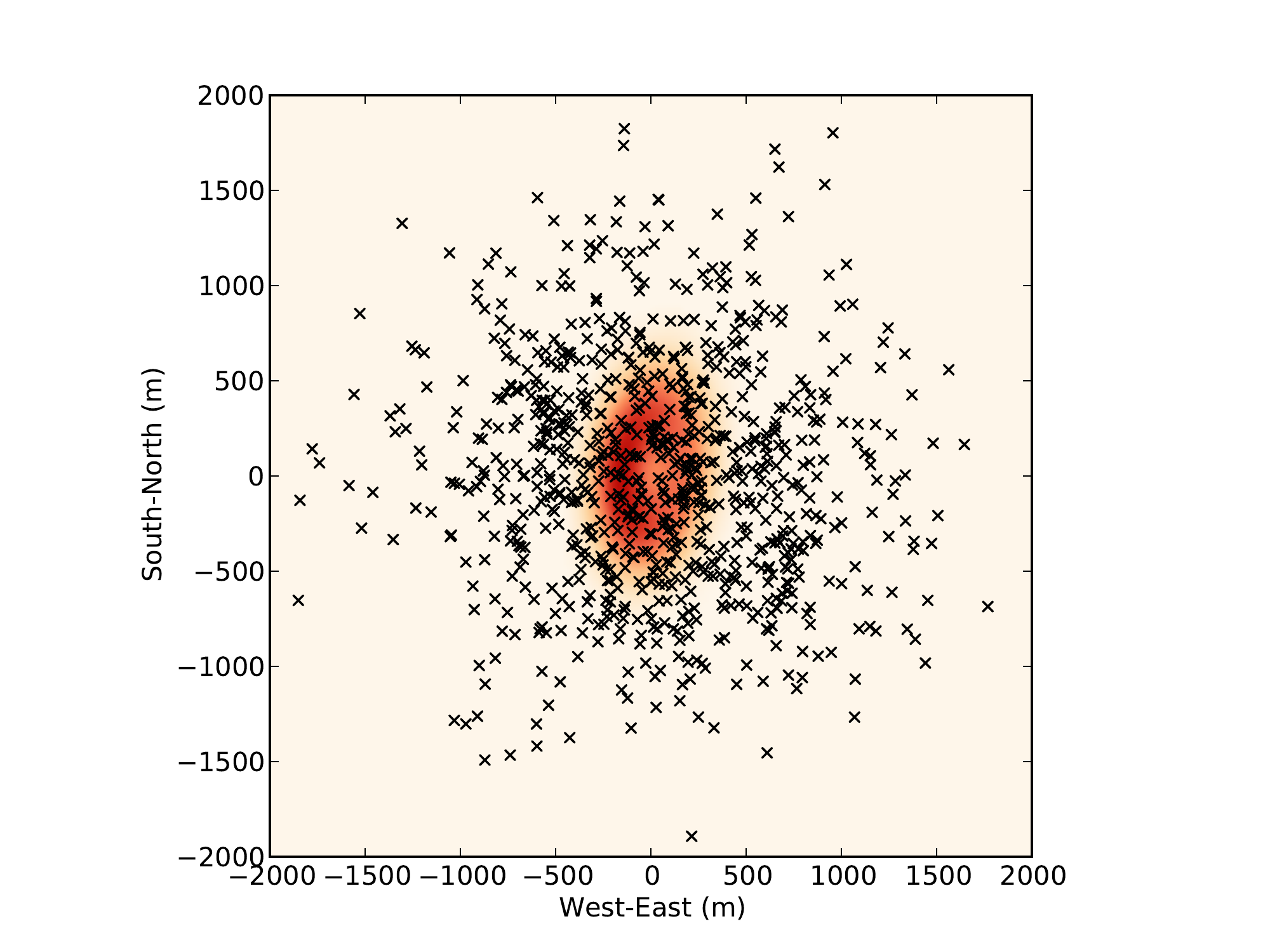}
  \caption{Antenna layouts for LOFAR (left), AERA (middle) and an 
  SKA-like array (right). The axes denote distances in metres.
  The background colors represent the radio footprint of an air shower 
  simulated with CoREAS. \citep{HuegeSKA2014}}\label{fig:ska}
\end{figure}

The second path being followed is that of very dense arrays such as 
LOFAR, which already today allows precise measurements of individual 
air showers. A new level in precision studies of air showers will be 
reached with the Square Kilometre Array (SKA) which is going to be 
available as of 2020 \citep{HuegeSKA2014}. The SKA will have a significantly larger area 
than LOFAR covered with a much more uniform array of antennas (Fig.\ 
\ref{fig:ska}), also 
spanning a larger bandwidth of 50--350~MHz. With these advantages, the 
SKA will be a precision instrument for cosmic ray studies that can be 
used to study the transition from Galactic to extragalactic cosmic 
rays, particle physics at energies beyond the reach of the LHC and 
possible connections between lightning initiation and cosmic rays.

\section{Conclusions}

The radio detection technique has matured within the previous decade. 
We understand the radio emission physics and have developed analysis 
strategies to extract the particle arrival direction, energy, and 
mass composition information. In the next few years, these will be 
evaluated precisely in direct comparison with established particle, 
fluorescence light and Cherenkov light detectors. The radio measurements 
will also provide benefits to the existing detectors, in particular a 
precise and independent cross-check of the absolute energy scale. 
Further strong potential for the radio technique exists in the large-scale application  
for inclined air showers and in precision studies with dense arrays 
such as LOFAR and the future SKA.





\bibliographystyle{aipproc}   

\end{document}